\documentclass[11pt]{article}

\usepackage[]{acl}

\usepackage{times}
\usepackage{latexsym}

\usepackage[T1]{fontenc}

\usepackage[utf8]{inputenc}

\usepackage{microtype}

\usepackage{pbox}
\usepackage{graphicx}

\graphicspath{ {images/} }
\usepackage{amsmath}

\title{End-to-End Document Classification and Key Information Extraction using Assignment Optimization}

\author{Ciaran Cooney \qquad Joana Cavadas \qquad Liam Madigan \qquad Bradley Savage \\ \qquad \textbf{Rachel Heyburn} \qquad \textbf{Mairead O'Cuinn}\\
 Aflac Northern Ireland \\
\texttt{\{ccooney,jcavadas,lmadigan,bsavage,} \\ \texttt{rheyburn,mocuinn\}@aflac.com}}

\begin{document}
\maketitle

\begin{abstract}
We propose end-to-end document classification and key information extraction (KIE) for automating document processing in forms. Through accurate document classification we harness known information from templates to enhance KIE from forms. We use text and layout encoding with a cosine similarity measure to classify visually-similar documents. We then demonstrate a novel application of mixed integer programming by using assignment optimization to extract key information from documents.  Our approach is validated on an in-house dataset of noisy scanned forms. The best performing document classification approach achieved 0.97 f1 score. A mean f1 score of 0.94 for the KIE task suggests there is significant potential in applying optimization techniques. Abation results show that the method relies on document preprocessing techniques to mitigate Type II errors and achieve optimal performance. 

\end{abstract}

\section{Introduction}
For many organizations, the ubiquity of personal computers and smart devices has led to the digitization of processes that previously required human involvement. In particular, paper documents are being digitized more frequently to enable electronic processing. Among those being digitized, documents such as invoices or insurance forms are common in daily workflows but often require tedious and costly manual processing or suffer from brittle automation systems \cite{majumder2020representation}. Automating workflows through rule-based systems or machine learning techniques can reduce the requirement for employees to engage in time-consuming data-entry and archiving tasks while cutting costs for employers \cite{audebert2019multimodal, mandivarapu2021efficient}. Document classification and key information extraction (KIE) are important tasks in automated document processing, and the two are not mutually exclusive. Classification of document images is often an important first stage in enterprise document processing upon which more granular downstream processing rests \cite{zhang2020text,appalaraju2021docformer,audebert2019multimodal,bakkali2020visual, kumar2014structural, harley2015evaluation, noce2016embedded}. KIE is one of those downstream tasks with the goal of automating retrieval of key information from documents \cite{majumder2020representation, gao2021field, gao2021value,schuster2013intellix, chiticariu2013rule, palm2019attend, garncarek2002lambert}.

Document image classification has been studied for decades \cite{peng2003document, sarkar2010learning,shin2001classification,taylor1995classification}. Early approaches considered functional landmarks and visual features of the spatial layout as discriminating factors between documents \cite{taylor1995classification,hu1999document,shin2001classification}. This strategy treats document images holistically and has been successful in classifying broad categories of documents \cite{taylor1995classification,shin2001classification,shin2006document}, but is less discriminating when applied to structurally- and visually-similar form-like documents \cite{harley2015evaluation}. For these, templating techniques have sometimes been employed \cite{peng2001document,peng2003document,sarkar2010learning}. Templating can be defined as finding the highest similarity between any document and a predefined set of document templates \cite{peng2003document}.

However, following computer vision-inspired developments in deep learning resulting from AlexNet \cite{krizhevsky2012imagenet} and ImageNet \cite{deng2009imagenet}, many approaches essentially treated document classification as a deep learning image classification problem \cite{afzal2015deepdocclassifier,kumar2012learning,kumar2014structural,kang2014convolutional,tensmeyer2017analysis,kolsch2017real}. Document images enable extraction of features such as graphics, typeface and colour, not possible with text-only approaches. Convolutional Neural Networks (CNN) were the primary driver of this development, with one of the initial applications being a 4-layer CNN used to classify tax forms and the Small Tobacco Dataset \cite{kang2014convolutional}. Other works used ImageNet pre-training as a starting-point for CNN-based classifiers \cite{harley2015evaluation,afzal2015deepdocclassifier}. However, \citet{tensmeyer2017analysis} argue that domain differences between natural images and documents limit the efficacy of this approach. In scenarios where broad document categories with distinct visual styles need to be classified, general features of each group positively assist classification (e.g., scientific papers differ greatly from magazine articles). However, within-group documents (e.g., forms from within a company - the category being investigated in this work) can exhibit high visual similarity and therefore require more granularity for accurate classification.

Increased granularity has been provided by a recent upsurge in multimodal approaches to document classification. Methods include combining textual representations with visual features \cite{noce2016embedded,bakkali2020visual,bakkali2022vlcdoc,audebert2019multimodal}, combining textual representations with positional embeddings \cite{xu2020layoutlm,wang2022lilt,li2021structurallm} or combinations of all three \cite{appalaraju2021docformer,xu2021layoutlmv2,huang2022layoutlmv3,gu2021unidoc, li2021selfdoc,powalski2021going}. \citet{audebert2019multimodal} demonstrated that their multimodal approach improves upon two unimodal baselines, while a dual-stream document classification technique combined word embeddings and visual features, with late fusion used to learn joint representations of documents \cite{bakkali2020visual}. 

LayoutLM, a large pre-trained model for document analysis, models interactions between text and layout information, with and without visual representations \cite{xu2020layoutlm}. At the time of publication, this approach achieved SOTA results in document image classification when combining all three modalities. Inspired by this, StructuralLM uses cell level 2d positional embeddings and token embeddings and a novel cell position training objective to achieve 96.08\% on RVL-CDIP \cite{li2021structurallm}. Another text and layout approach, LiLT, uses bi-directional attention to model cross-modal interactions to achieve similar performance (95.68\%) \cite{wang2022lilt}.

 Other general document analysis approaches, such as LayoutLMv2 and LayoutLMv3 have adapted the way visual features are incorporated in multimodal approaches, including using masked image modelling and word-patch alignment training objectives to ensure cross modal alignment \cite{xu2021layoutlmv2,huang2022layoutlmv3}. Despite having proven that visual features such as text font and visual layout are useful for extracting general differences between documents, it is not clear that they perform as well when differentiating between very similar documents, a common problem in industry applications. Additionally, these large pre-trained models require significant computing resources and large training datasets which are not always available. We address some of these issues by using a classification approach that relies on document templates rather than large-scale training.

Often downstream of document classification, KIE, including value retrieval, involves the extraction of key information from structured documents such as forms or invoices. Extracted information can be used for many tasks including customer enrollment and insurance claim adjudication. Several early approaches to KIE relied on templates constructed to enable cross-referencing of new documents against a bank of existing templates \cite{rusinol2013field, chiticariu2013rule}. \citet{rusinol2013field} developed user-generated training samples which they used to build a document model based on structural templates. Other approaches relied on pre-registered templates within the system to perform KIE \cite{chiticariu2013rule, schuster2013intellix}. However, \citet{chiticariu2013rule} showed a disconnect between industry and academia w.r.t. the utility of rules-based approaches to IE. Although industry applications often require some form of rules-based intervention in production systems, template and rule-based methods are constrained to specific layouts and may not generalize well to unseen documents.

In the era of deep learning it is common to formulate KIE as a sequence-labelling task \cite{huang2015bidirectional, lample2016neural}. However, this approach does not handle complex spatial relationships and is not ideal for highly structured documents \cite{hwang2021spatial}. Due to this there have been several developments in deep learning approaches that have gone beyond the sequence-labelling formulation. \citet{palm2019attend} proposed a CNN approach to KIE that rejected the need for word-level labels, and therefore is useful in real-world scenarios where labelled data is not always available. Another novel approach, Doc2Dict is a T5 transformer trained on database records to produce a document-to-data-structure model \citet{townsend2021doc2dict}. \citet{zhang2020text} proposed an end-to-end text reading and KIE method that extracts entity values through a multimodal fusion of text and visual features, and \citet{majumder2020representation} presented a field-value pairing framework that utilizes knowledge about the data-types of fields to select candidate entities. In the latter, a set of candidates is identified as possibly corresponding to a field in the target schema. A neural network is used to learn representations of each candidate based on neighbouring words, before selecting a correct value. This approach is useful in scenarios with high volumes of unseen documents. 

As the structure of forms is often more sophisticated than a simple linear ordering of tokens, relative token position, paragraph spacing, and question-answer relationships all contain information that can be leveraged for KIE tasks \cite{garncarek2002lambert}. Chargrid \cite{katti2018chargrid} , CUTIE \cite{zhao2019cutie} and BERTgrid \cite{denk2019bertgrid}  were among the first models to integrate 2D representations of word tokens alongside text for KIE, and each outperformed their respective baselines. Utilizing the 2D positions of text to assist KIE has helped architectures such as LAMBERT \cite{garncarek2002lambert} and the LayoutLM family \cite{xu2020layoutlm, xu2021layoutlmv2, huang2022layoutlmv3} achieve SOTA performance while exhibiting less sensitivity to serialization. Whereas LayoutLM uses tokens, layout and visual information, LAMBERT relies only on tokens and bounding boxes.

This work introduces an end-to-end document classification and KIE pipeline based on a templating approach that does not require any model training.We eschew the recent trend towards deep learning focused approaches to document classification and KIE while retaining the important consideration of 2D document structural layout.We demonstrate that different text encoding strategies work remarkably well for document classification when computing cosine similarity between a document and a set of document templates. A novel assignment optimization technique for KIE is presented which assigns values in a form to a corresponding template key based on global geometric positions and specified constraints. This approach is insensitive to serialization and word tokenization and does not require large training data while being easily updated. We report a series of processing steps that are required to make assignment optimization feasible for noisy scanned documents, and test these with ablation experiments. Finally, we detail some limitations of our approach and suggest future directions for refining the technique further.

This work makes the following contributions: 
\begin{itemize}
  \item We present a novel assignment optimization approach to key information extraction.
  \item We present a granular document classification strategy by finding the cosine similarity between a vectorized document and a matrix of document templates.
  \item We demonstrate the importance of document rotation and entitiy scaling processing steps in maximising the potential of our optimization approach.
\end{itemize}

\section{Related Work}
\subsection{Document Classification}
Features can be extracted from document images to represent either text content or visual properties. Image templates have often been favoured in industrial document analysis \cite{sarkar2010learning}. \citet{sarkar2010learning} use image anchor templates for document classification and propose a method for learning templates from few training examples. Combining both text and visual content is a popular approach to document classification. \citet{noce2016embedded} reported performance improvement when supplementing visual features with text, especially in cases where different classes share visual characteristics. StructuralLM \cite{li2021structurallm} and LiLT \cite{wang2022lilt} are deep learning approaches that demonstrated the value of combining text and layout information for document classification.
With the proliferation of text encoding techniques such as Word2Vec \cite{mikolov2013efficient} and ELMO \cite{sarzynska2021detecting} several studies have used off-the-shelf algorithms to generate representations of document images for classification \cite{bakkali2020visual, audebert2019multimodal}.

\subsection{Key Information Extraction}

KIE has received focused attention, with several early approaches relying on templates constructed to enable cross-referencing of new documents against a bank of existing templates \cite{rusinol2013field, chiticariu2013rule, schuster2013intellix}. Recently, deep learning approaches have dominated the literature. \citet{palm2019attend} present a deep neural network that bypasses the need for word-level training labels, with the assumption that the same structured information must be extracted from each document. Another approach seeks to predict target values from arbitrary queries \cite{gao2021field}. The technique utilises a novel pre-training strategy which makes it more flexible for learning local geometric relations between words. \citet{gao2021value} also propose a novel framework for using unlabelled documents for training with known form types. Attempting to overcome expensive annotation efforts, the approach uses a bootstrapped training process to develop a rule-based approach for getting pseudo-labels from unlabelled forms. Pseudo-labels are then used for supervised learning. Related to this approach for reducing annotation effort, findings that bounding boxes alone can be effective in VrDU tasks 
\cite{cooney2023unimodal}, and work formulating KIE as a constrained optimization problem through partial graph matching \cite{yao2021one}, we propose an assignment optimization approach to KIE.

\subsection{Assignment Optimization}
An assignment problem has a number of \emph{agents} and a number of \emph{tasks}. Agents are assigned to perform tasks, with a cost depending on agent-task assignment. \citet{gong2021transfer} optimize passenger-route assignment by formulating it as a nonlinear mixed integer optimization model. Other problems such as resource allocation can be formulated as assignment problems to assign a number of tasks (e.g. jobs), to a number of workers (agents), to maximise or minimise some utility function \cite{lee2018deep}. In our formulation, agents are template value-positions corresponding to a specific key. Tasks are form entities. The cost we seek to minimize is the Euclidian distance between template value-positions and form entities.

\section{Dataset} \label{dataset}

\begin{figure*}[t]
\centering
\includegraphics[width=\textwidth]{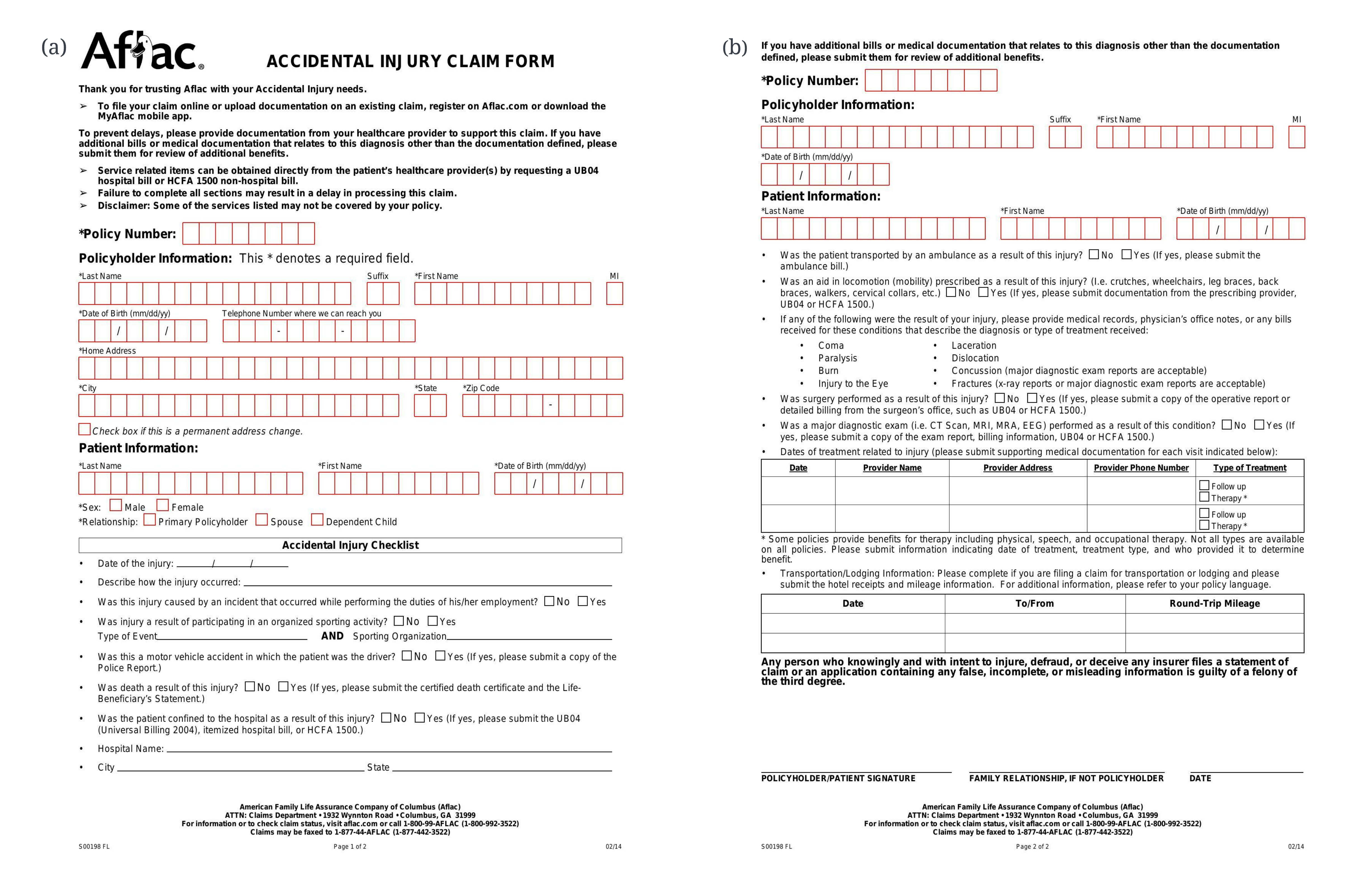}
\caption{Two of the six forms in our template dataset. (a) Aflac Accidental Injury Claim Form (page 1). (b) Aflac Accidental Injury Claim Form (page 2). Forms consist of keys and value spaces to be filled-in by claimants.}
\label{fig:templates}
\end{figure*}

Datasets used in document classification studies typically  consist of broad categories of documents (e.g. email, letter, scientific report) \cite{harley2015evaluation, kumar2014structural}. Our data consists of 395 scanned document images, each corresponding to one of six categories of health insurance claim forms and exhibiting a degree of structural homogeneity common within organizations (Figure \ref{fig:templates}). Each form contains a set of \emph{keys} adjacent to an associated text-box or white-space into which information (\emph{values}) can be entered. Each key in a form asks a claimant to enter a specific piece of information (e.g. first name, last name, policy number), but not all value spaces have been filled-in. Forms in the dataset contain both printed and handwritten responses to key requests for information. The scanned documents are of fixed dimensions (1700 × 2200 pixels) and  exhibit significant rotation variance and noise. The number of key-value pairs per document class ranges from 217 to 2484. Dataset statistics are reported in Table \ref{tab:dataset}.

\begin{table}
\small
\centering
\begin{tabular}{lll}
\hline
\pbox{1cm}{\textbf{Form}} & \pbox{2cm}{\textbf{N documents}} & \pbox{2cm}{\textbf{N key-values}}  \\
\hline
\verb|aicf_pg1| & {} {} {} {} {} {} {169} & {} {} {} {} {} {} {2484} \\
\verb|aicf_pg2| & {} {} {} {} {} {} {144} & {} {} {} {} {} {} {1390} \\
\verb|hicf_pg1|  & {} {} {} {} {} {} {30} & {} {} {} {} {} {} {422} \\
\verb|hicf_pg2| & {} {} {} {} {} {} {24} & {} {} {} {} {} {} {352} \\ 
\verb|aicf_v1| & {} {} {} {} {} {} {14} & {} {} {} {} {} {} {217} \\
\verb|pvbcf| & {} {} {} {} {} {} {14} & {} {} {} {} {} {} {279} \\
\hline
\end{tabular}
\caption{Dataset statistics for each form type.}
\label{tab:dataset}
\end{table}

We use Amazon Textract\footnote{\url{https://aws.amazon.com/textract/}} Optical Character Recognition (OCR) engine to extract text and bounding boxes from the scanned documents and the six document templates.

\subsection{Consolidating OCR Output}
\begin{figure}[t]
  \centering
  \includegraphics[width=\linewidth]{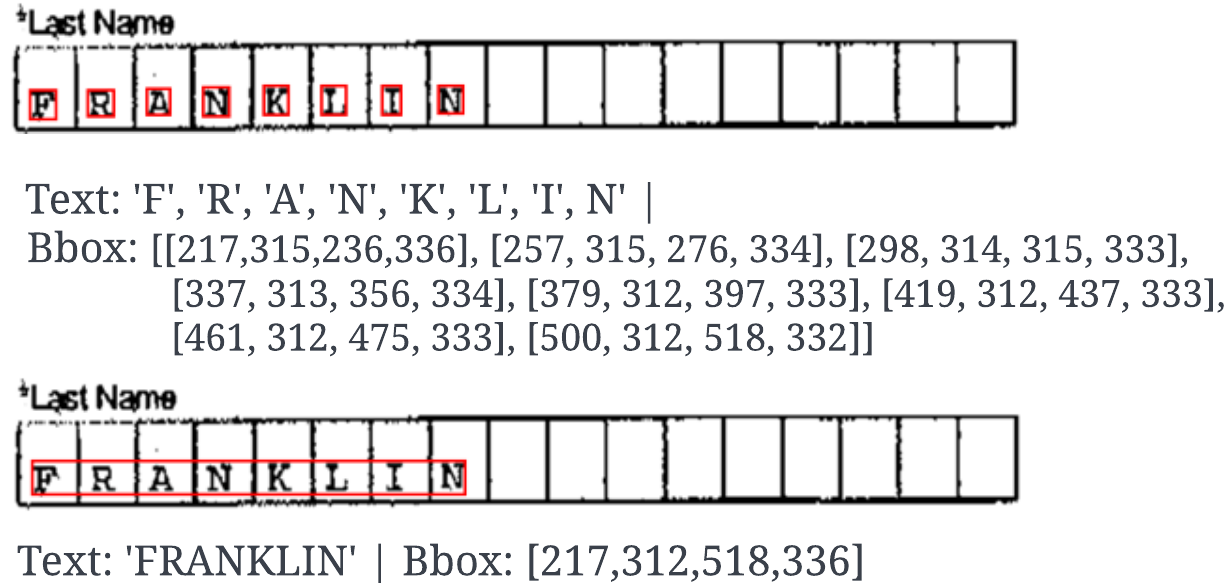}
  \caption{Example of original character-by-character output from OCR (top), and consolidated entity (bottom).}
  \label{fig:ocr_consolidation}
\end{figure}

Previous studies have noted the negative impact that OCR mistakes can have on KIE tasks \cite{audebert2019multimodal, palm2019attend}. Due to the noisy nature of many of our scanned documents, OCR output can be degraded in terms of identifying characters that form part of an entity string (Figure~\ref{fig:ocr_consolidation} - top). To deal with this, we implement a simple yet effective method for consolidating character strings that are likely to form part of a single entity. First, we iterate through the set of strings identified by OCR from top-left to bottom-right of the document. For each source string (i.e., the string we may wish to append to), we search potential candidate strings to determine whether they are vertically aligned with the source. Here, we use a threshold of ±15 pixels for both top and bottom of the bounding boxes. If bounding boxes are inline, we then compute the distance along the horizontal axis between the trailing character in the source string and the leading character in the candidate string. If this distance is below a specified threshold, the character strings are consolidated, and their bounding boxes merged (Figure~\ref{fig:ocr_consolidation} - bottom). If the difference between the start of a source and start of a target is <60 pixels we append as a single word. If >60 we append as if separate words within an entity (e.g. parts of an address).

\subsection{Document Image Alignment} \label{homography}
KIE from document images can suffer from geometrical distortions caused by scanning processes \cite{narayan2017document}. To mitigate the effects of rotation variance, we implement document image alignment to rotate scanned documents to align with their corresponding template images. First, we use Oriented FAST and rotated BRIEF (ORB) to detect keypoints within a document and to extract local invariant descriptors \cite{rublee2011orb}. Then, applying Hamming distance to compute distances between template and scanned features, we determine a set of best matches. Next, random sample consensus (RANSAC) matches keypoints between the template and the scanned  document \cite{fischler1981random}, before computing a homography matrix between the two documents. The homography matrix is a perspective transformation between the template and scanned document, facilitating alignment of the scanned document. This method was implemented in OpenCV \cite{bradski2000opencv}.  

\section{Methods}


\subsection{Global Representations for Document Classification}
As stated in the introduction, documents from different classes that exhibit similar features require fine-grained analysis rather than broad visual features to enable differentiation.  
Here, document-level templates are constructed from vectorized document text and layout embeddings generated from bounding box information returned from the OCR. We implemented three different methods for vectorizing the document text: two versions of the Universal Sentence Encoder (USE) \cite{cer2018universal} and Term Frequency Inverse Document Frequency\footnote{\url{https://scikit-learn.org/stable/modules/generated/sklearn.feature_extraction.text.TfidfVectorizer.html}} (TF-IDF). 

The USE approaches are designed to be general purpose text embedding models. Both USE models receive input English strings and return a fixed 512 element vector representation of the string as a sentence embedding, which here represents the entire document. The first USE method is a transformer-based encoding model which generates text embeddings by using attention to compute context aware representations of words \cite{vaswani2017attention}. The second is a deep averaging network which computes average input embeddings for words and bi-grams before being passed through a feedforward deep neural network to obtain a 512 element sentence embedding.

TF-IDF computes the relative frequency of words in one document compared to the inverse frequency of that word in the entire corpus. In this case the document corpus consists of the combined text from the six forms (Section \ref{dataset}). The TF-IDF algorithm is fitted to the data to learn its vocabulary and inverse document frequency. Each of the six forms is then transformed into a document-term matrix using the fitted vectorizer. The fitted vectorizer is retained to transform new unseen forms.

\begin{figure}[t]
  \centering
  \includegraphics[width=\linewidth]{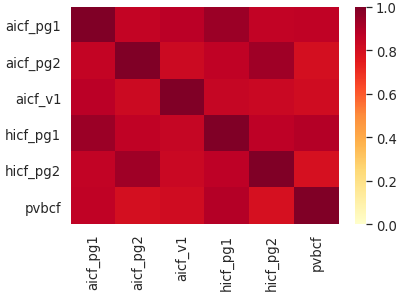}
  \caption{Correlation between embedded text representations of document templates is high.}
  \label{fig:template_correlation}
\end{figure}

Forms often exhibit similar structure and when taken from a single domain can often contain duplication of text (e.g., institutional logos, requests for similar information). Figure~\ref{fig:template_correlation} depicts the correlation between USE vector representations of the template documents, indicating the difficulty in distinguishing between forms using this approach. In particular, forms \emph{aicf\char`_pg1} and  \emph{hicf\char`_pg1} (0.943) and forms \emph{aicf\char`_pg2} and  \emph{hicf\char`_pg2} (0.934) are highly correlated. Here we supplement our text embeddings by also generating layout embeddings for each form. These embeddings represent the 2D positions of the text and are generated using the pretrained Layoutlmv2 model \cite{xu2021layoutlmv2}.  

\subsection{Document Classification using Cosine Similarity}

Document class is predicted by measuring the cosine similarity between a new document and a set of document templates. A template matrix \(u\) is constructed as a \(n \times m\) matrix where \(n\) is the number of templates in the template bank and \(m\) is the length of the vector representation of a single document. Each new form representation to be measured against the template matrix is represented as a \(1 \times m\) vector \(v\). Cosine distance between vectors \(u\) and \(v\) is formulated as:
\begin{equation} \label{eq1}
s(u,v)  = 1- \frac{u \cdot v}{||u||_2||v||_2}
\end{equation}
\newline
\emph{where}, \({||{*}||_2} \) is the 2-norm of argument {*}, and \(u \cdot v\) is the dot product of \(u\) and \(v\). This returns a \(n \times 1\) matrix containing similarity between the candidate form \emph{v} and \emph{n} templates in matrix \emph{u}. We then take the \emph{argmax} of the similarity matrix to obtain a document classification.

\subsection{Problem Formulation for KIE}

Document classification is followed by KIE, where templates are constructed for each of the six document classes in our dataset. The goal of KIE is to identify entity values in a document that are associated with a specific key. Given a document template:

\begin{multline*}
DT  = \Bigl\{\Bigl(kw^{(i)}, vx_{min}^{(i)}, vy_{min}^{(i)}, vx_{max}^{(i)}, vx_{max}^{(i)}\Bigr)\\ {|\space} \;i \in \{1,...,n\}\Bigr\}                      
\end{multline*}

consisting of \emph{n} key words \emph{kw\textsuperscript{(i)}} and a bounding box associated with their corresponding value positions $vx_{min}^{(i)}$, $vy_{min}^{(i)}$, $vx_{max}^{(i)}$, $vy_{max}^{(i)}$, and a document:

\begin{multline*}
D  = \Bigl\{\Bigl(w^{(j)}, x_{min}^{(j)}, y_{min}^{(j)}, x_{max}^{(j)}, x_{max}^{(j)}\Bigr)\\ |\space \;j \in \{1,...,m\}\Bigr\}                      
\end{multline*}
consisting of \emph{m} words \emph{w\textsuperscript{(j)}} and their associated bounding box $x_{min}^{(j)}$, $y_{min}^{(j)}$, $x_{max}^{(j)}$, $y_{max}^{(j)}$, the objective is to correctly assign \emph{n} document bounding boxes to each of the template bounding boxes. In doing so, we assign associated words (values) to the set of key words in the template, thus producing key-value pairs. 

Each KIE template consists of a set of key-value pairs. Template keys are text (e.g. 'Name', 'D.O.B.'), and values are a bounding box corresponding to a location on the page where information can be entered. 
With our method of applying assignment optimization to the KIE task, each bounding box is represented by a single coordinate point. In initial experiments, the bounding box centroid was selected as the 2D coordinate to represent the position of each entity. However, this approach proved unstable during optimization as differences in length between unfilled template values and unseen form values routinely resulted in unexpected mis-assignments. Our reported results are based on selection of the upper-left point of each bounding box representing an entity position, as used by \citet{katti2018chargrid}. The reason for this is that while the length of form value entries vary significantly and therefore skew the centroid position, the starting position of an entity value is relatively fixed, providing a more stable representation.


\subsection{Scaling Document Entities} \label{scaling}

\begin{figure*}[t]
\centering
\includegraphics[width=\textwidth]{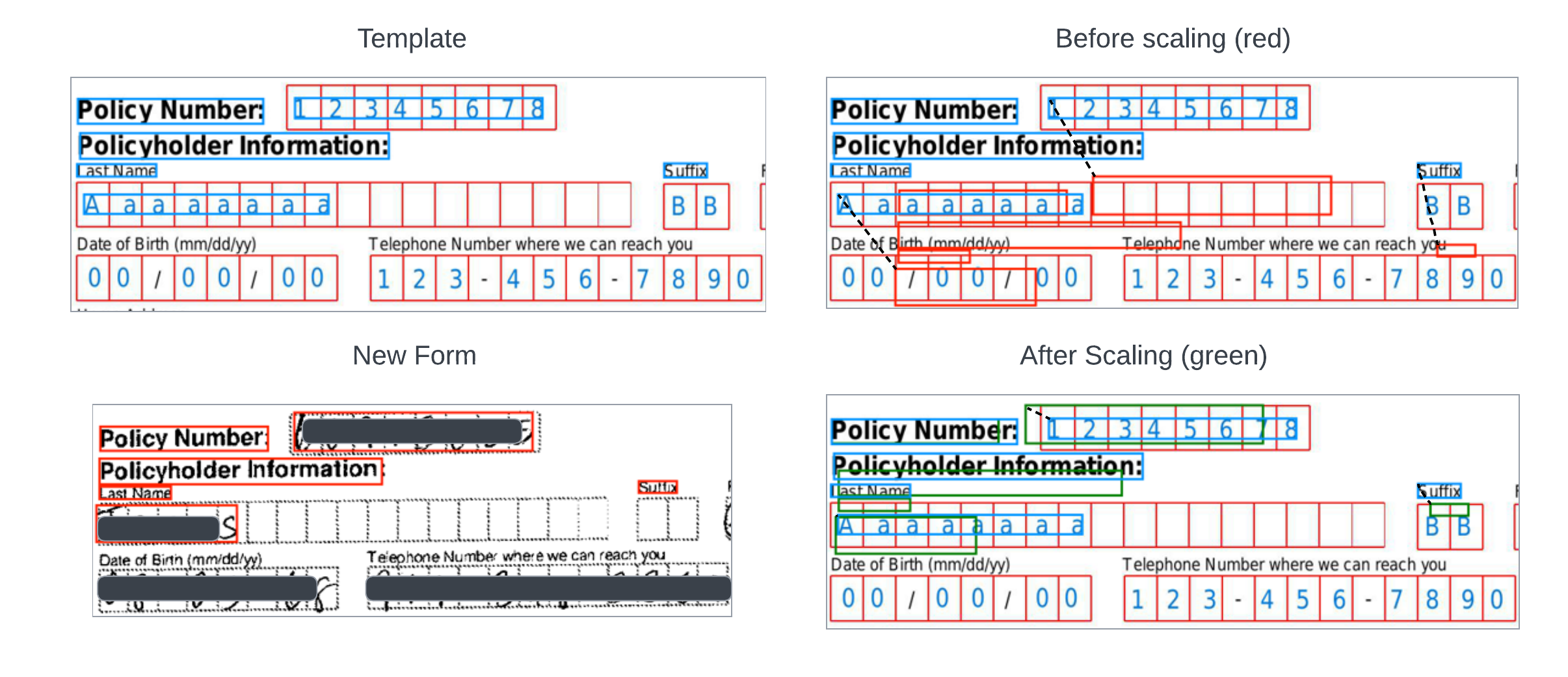}
\caption{Entities in a scanned document form are scaled to more closely approximate the assigned template. Red bounding boxes indicate scanned document bounding box positions pre-scaling. Green boxes indicate positions post-scaling.}
\label{fig:doc_scaling}
\end{figure*}

A virtue of using accurate document classification to select templates for KIE is that prior information on the specific form being processed can be used to enhance optimization. An issue with scanned documents is significant variation in scale, rotation, and aspect ratio in comparison with the original template versions of these forms \cite{ahmad2021efficient}. Here, we use prior knowledge from form templates to scale unseen form entities to approximate dimensions of the template (Figure \ref{fig:doc_scaling}).  

As a first step, we compute coordinates for twenty rectangular segments of equal area. This segmentation is required because scanned documents are not always warped linearly e.g., entities at the top of a page can be offset by a different factor to those at the bottom. Then, within each segment we search for text strings extracted by OCR. Next, we check whether a word within a given segment is present among the set of keys in our entity template. For this, we implement fuzzy matching with a minimum confidence level of 0.9 to account for errors in OCR outputs. To avoid mis-scaling due to duplicate terms within a document, we apply a maximum distance threshold. 

Within each segment, Manhattan Distance is measured between each text string and a matching template keyword:
\begin{equation} \label{eq2}
d(x,y)  = |x_1 - x_2| + |y_1 - y_2|
\end{equation}
\emph{where}, (\(x_1, y_1\)) and (\(x_2, y_2\)) are 2d coordinates of the top-left of template and new form entity bounding boxes, respectively.
Mean scalar values for each axis are calculated for each segment independently, resulting in distinct vertical and horizontal scalars for scaling each segment.
Using these scalars, we re-scale the unseen form entities to approximate positions in the template (Figure~\ref{fig:doc_scaling}).

\subsection{Assignment Optimization}

To assign entities from a new form to keys in a template, we developed a binary optimization model. This approach eliminates any requirement for data-hungry neural network training. The optimization objective is to minimise the distance between entities in a new form and spaces in a form template where values are entered i.e., when a text string in a form has been entered into one of the template value positions, we expect to see a close-to-zero distance between the two bounding boxes. Of course, text in a form that does not correspond to some filled-in value is expected to have a relatively high distance to our template bounding boxes.

Let $i \in T_k$ and $j \in F_l$ be a value-position within a template $T_k \in \mathcal{T}$ and an entity within the form $F_l \in \mathcal{F}$, respectively. The standard euclidean distance $D_{ij}$ between $i$ and $j$ is defined as (\ref{eq: distance}), with $\mathcal{T}_i^k$ and $\mathcal{F}_j^l$ being the coordinates of value-position and entity bounding boxes, respectively.

\begin{equation}
D_{ij} = \mathcal{T}_i - \mathcal{F}_j  , i \in T, j \in F
\label{eq: distance}
\end{equation}



The optimization model used to assign form entities to template value-positions comprises the following features:

\begin{itemize}
    \item $\mathcal{T}$: form template.
    \item $\mathcal{T}_i = (t_i^x, t_i^y)$: coordinates of value-position $i$ within template (top-left corner).
    \item $\mathcal{F}$: form.
    \item $\mathcal{F}_j = (f_j^x, f_j^y)$: coordinates of entity $j$ within form (top-left corner). 
    \item $\mathcal{D}_{ij}$ : distance between value-position $i$ from template  value-position $T_i$ and entity $j$ from form entity $F_j$.
    \item $\mathcal{M}_{ij}$ : set of form entities matching template keys.
\end{itemize}

\begin{align}
\min & \sum_{i \in \mathcal{T}} \sum_{j \in F} \mathcal{D}_{ij} \cdot x_{ij} \label{opt: ObjectiveFunction}\\
\textrm{s.to} &  \sum_{j \in \mathcal{F}}  x_{ij} = 1, \hspace{2mm} i \in \mathcal{T} \label{opt: Cond1}\\
              & \sum_{i \in \mathcal{T}}  x_{ij} \leq 1, \hspace{2mm} j \in F\label{opt: Cond2}\\
              & x_{ij} = M_{ij}, \hspace{2mm} i \in \mathcal{T}, j \in F\label{opt: Cond3}, \\ 
              & x_{ij} \in \{0,1\} \textrm{ for all } t  \textrm{ in } T \textrm{ and all } f \textrm{ in } F \label{opt: DecisionVariableDomain}
\end{align}

where (\ref{opt: ObjectiveFunction}) tries to minimize global distance between template value bounding boxes and new form bounding boxes by assigning a maximum of one form bounding box to one template bounding box. The final output is a set of pairs $x_{ij}$, where \emph{i} is a template bounding box and \emph{j} is a form bounding box, such that $x_{ij}=1$. 

\begin{figure}[t]
  \centering
  \includegraphics[width=\linewidth]{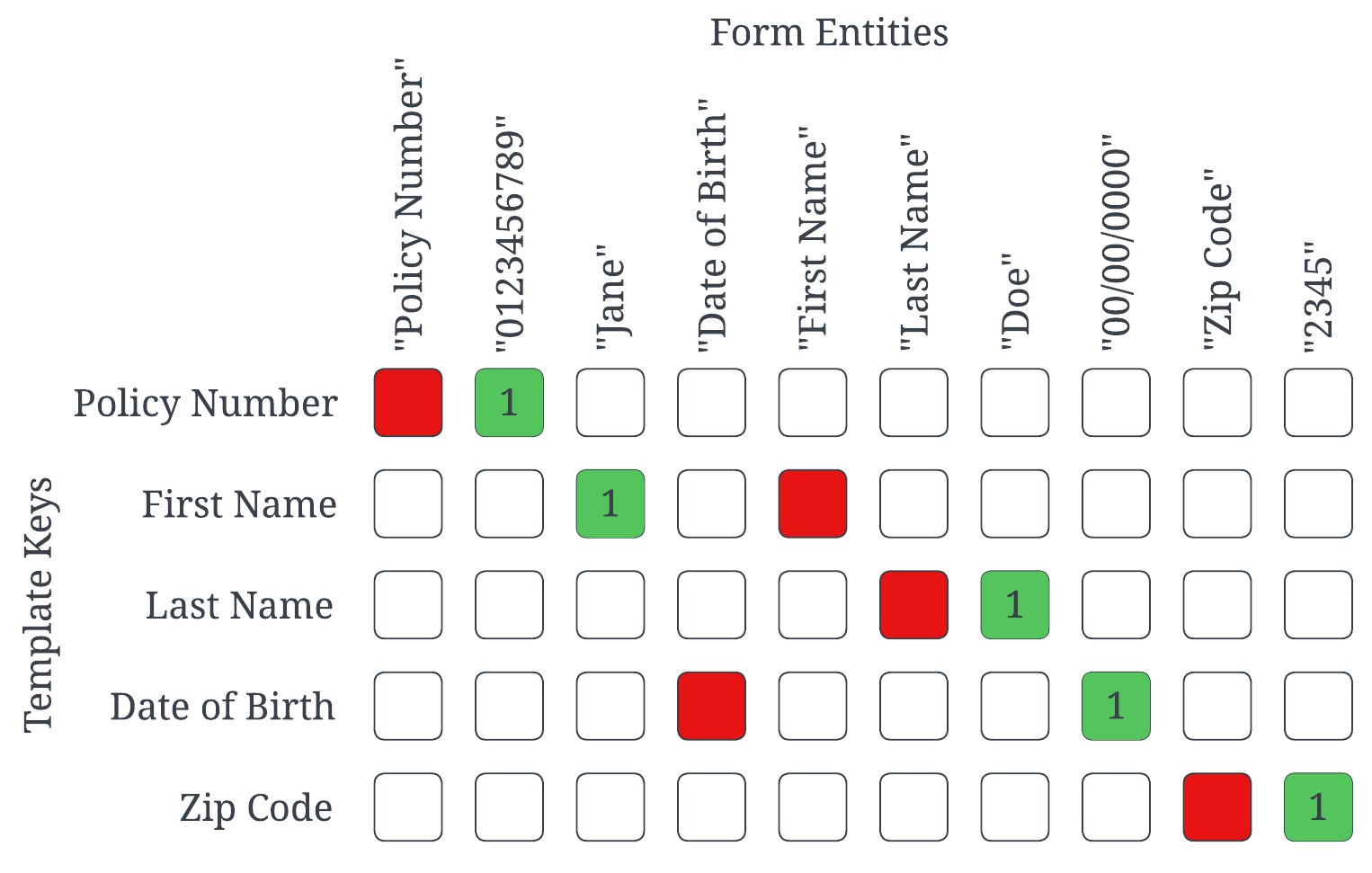}
  \caption{Assignment optimization for form entities to template keys. Rows are template positions, columns are new form positions. Red squares are constraints indicating a columns can't be assigned to a row. Green squares indicate assignment.}
  \label{fig:assigment_optimization}
\end{figure}

Figure \ref{fig:assigment_optimization} depicts the implementation of assignment optimization for the KIE task. Bounding-boxes for template key-value positions (e.g. 'Policy Number', 'Last Name') are designated rows, while form entity (e.g. 'Doe', '0123456789') bounding boxes are designated columns. Our optimization approach assigns entities to keys by minimizing overall distance between form entities and template value-positions, with constraints. Green squares indicate that a form entity has been assigned to a template key. Red indicates a constraint on assigning certain entities to certain keys. White squares indicate that an assignment is possible but has not been made for this solution.



\section{Results}
The transformer-based USE using text and layout encodings achieved a weighted average f1 score of 0.97 for the document classification task. In total, 383 out of 395 forms were assigned to the correct template label. The deep random network and TF-IDF approaches also performed well with f1 scores of 0.95 and 0.94, respectively. Form \emph{hicf\char`_pg2} accounts for the majority of misclassified instances, conforming with similarity scores between templates (Figure~\ref{fig:template_correlation}).

Precision, recall, and f1 scores are used to evaluate the performance of our assignment optimization approach to KIE (Table~\ref{tab:ie results}). With a mean f1 score of 0.941, our method exhibits strong performance for the KIE task despite the noise and rotation variance present in our dataset. Mean precision and recall scores of 0.954 and 0.928 suggest that the current implementation of the approach is slightly more likely to misclassify entities in the form than assign values to spaces that were actually unfilled. This means we can have a high degree of confidence that values are assigned to the correct keys, but less confidence that all entities have been assigned.

A striking result from Table~\ref{tab:ie results} is the difference in precision (0.962) and recall (0.872) for \verb|hicf_pg2|. The reason for this relatively poor recall score appears to be the density of text surrounding the key-value pairs in that particular form. The close proximity of extraneous text to values to be extracted, and the imperfect document scaling technique combine to produce a relatively large number of misassigments. Another notable result is the performance of our method at extracting information from the \verb|aicf_v1| form. The proposition that KIE algorithms perform more or less effectively depending on the type of document they face is not new \cite{zhao2019cutie,cheng2022trie++}.  

Results are comparable with other KIE techniques. On the SROIE dataset, LayoutLM, StructText, LayoutLMv2, TILT, LAMBERT achieved mean KIE f1 scores ranging from 0.952 - 0.982 \cite{xu2020layoutlm, li2021structext, xu2021layoutlmv2, powalski2021going, garncarek2002lambert}, and TRIE++ recently reported 0.984 \cite{cheng2022trie++}. Although f1 scores presented here are slightly inferior, our approach offers the benefit of very specific key labelling that can be important for industrial applications of IE.

\begin{table}
\small
\centering
\begin{tabular}{llll}
\hline

\pbox{2cm}{\textbf{Document}} & \pbox{2cm}{\textbf{Precision}} & \pbox{2cm}{\textbf{Recall}} & \pbox{2cm}{\textbf{F1}} \\
\hline
\verb|aicf_pg1| & {0.962 (2312/2404)} & {0.931 (2312/2484)} & {0.946}\\
\verb|aicf_pg2| & {0.950} (1304/1372) & {0.938} (1304/1390) & {0.944}\\
\verb|aicf_v1| & {0.887} (182/205) & {0.839} (182/217) & {0.862} \\ 
\verb|hicf_pg1| & {0.954 (393/412)} & {0.931 (393/422)} & {0.942} \\ 
\verb|hicf_pg2| & {0.962 (307/319)} & {0.872 (307/352)} & {0.915} \\
\verb|pvbcf| & {0.939 (275/293)} & {0.986 (275/279)} & {0.962}  \\ \hline
\pbox{2cm}{\textbf{Mean}} & 0.954 & 0.928 & 0.941
\end{tabular}
\caption{Precision (TP/(TP+FP)), Recall (TP/(TP+FN)) and F1 scores for each document type.}
\label{tab:ie results}
\end{table}

\subsection{Ablation Study}

To evaluate the impact of preprocessing steps on the overall performance of our KIE method we performed an ablation study. In separate experiments, we remove document image alignment (Section~\ref{homography}) and entity scaling (Section~\ref{scaling}) from the pipeline. Results in Table~\ref{tab:no homography results} demonstrate an almost 10\% drop in mean F1 score when documents are not aligned with template images. A similar decrease in performance is observed when entity scaling is removed from the pipeline (Table~\ref{tab:no scaling results}).

The degraded performance is largely due to an increase in Type II errors (false negatives), as witnessed by a sharp decline in recall scores. This vindicates the use of these preprocessing steps to augment our assignment optimization algorithm that would otherwise struggle with noisy and skewed scanned documents. Without rotation and scaling our algorithm fails to classify approximately 20\% of key entities we should extract from our forms.

\begin{table}
\small
\centering
\begin{tabular}{llll}
\hline

\pbox{2cm}{\textbf{Document}} & \pbox{2cm}{\textbf{Precision}} & \pbox{2cm}{\textbf{Recall}} & \pbox{2cm}{\textbf{F1}} \\
\hline
\verb|aicf_pg1| & {0.921} & {0.747} & {0.825}\\
\verb|aicf_pg2| & {0.952} & {0.791} & {0.864}\\
\verb|aicf_v1| & {0.894} & {0.734} & {0.806} \\ 
\verb|hicf_pg1| & {0.925} & {0.874} & {0.899} \\ 
\verb|hicf_pg2| & {0.948} & {0.786} & {0.859} \\
\verb|pvbcf| & {0.857} & {0.768} & {0.810}  \\ \hline
\pbox{2cm}{\textbf{Mean}} & 0.916 & 0.783 & 0.844
\end{tabular}
\caption{Precision, Recall and F1 when realignment is not applied.}
\label{tab:no homography results}
\end{table}

\begin{table}
\small
\centering
\begin{tabular}{llll}
\hline

\pbox{2cm}{\textbf{Document}} & \pbox{2cm}{\textbf{Precision}} & \pbox{2cm}{\textbf{Recall}} & \pbox{2cm}{\textbf{F1}} \\
\hline
\verb|aicf_pg1| & {0.923} & {0.745} & {0.825}\\
\verb|aicf_pg2| & {0.958} & {0.824} & {0.886}\\
\verb|aicf_v1| & {0.918} & {0.774} & {0.840} \\ 
\verb|hicf_pg1| & {0.922} & {0.839} & {0.879} \\ 
\verb|hicf_pg2| & {0.983} & {0.841} & {0.906} \\
\verb|pvbcf| & {0.909} & {0.860} & {0.883}  \\ \hline
\pbox{2cm}{\textbf{Mean}} & 0.936 & 0.788 & 0.856
\end{tabular}
\caption{Precision, Recall and F1 when document scaling is not applied.}
\label{tab:no scaling results}
\end{table}

\section{Limitations}

A limitation of this work is that it does not fully consider how to integrate extraction of information from tables using our optimization approach. Several recent studies have reported entity extraction from tables \cite{paliwal2019tablenet, yang2022tableformer, nazir2021hybridtabnet, agarwal2021cdec}, suggesting feasible approaches to this task. For example, \citet{katti2018chargrid} and \citet{denk2019bertgrid} tackle tabular data extraction by predicting the coordinates of the table rows bounding boxes to identify the invoiced products. Another limitation is that skew detection and document dewarping are not currently a fully automated process in our pipeline. Methods for skew detection and correction, such as those reported in \citet{ahmad2021efficient} could be incorporated in future work to aid processing. Our method has been tested on a limited number of form classes, and currently it does appear to be weaker when faced with densely-populated forms and documents. This is unsurprising, but methods for dealing with this are required to make this method generalise to a more diverse set of forms.

Another potential limitation of the system is that the Transformer-based USE is more expensive than the deep averaging network in terms of compute time \cite{cer2018universal}. The deep averaging network compute time is linear with length of string, but classification accuracy was not as strong as the transformer approach. Document classification based on similarity assigns a class even when a document is not one of the classes. In a production system, we would consider an 'other' category for such cases.

\section{Future Work}

Results indicate that our optimization approach to KIE is not entirely robust to all document structures. It will be important to experiment with techniques to improve the generalizability of this method to a more diverse range of documents. Reducing sensitivity to closely located text will help reduce some of the Type II errors. Additionally, it is possible that setting distance constraints between templates and new forms along two dimensions may enhance our assignment operation. 

Another area of future work will seek to improve the input to our optimizer. OCR engines can have difficulty dealing with obscure fonts, diacritics, or other relatively uncommon artefacts \cite{audebert2019multimodal}, and we know that the method proposed here relies on strong OCR performance to achieve reliable results. One of the areas to improve our approach is to enhance the fidelity of the initial text recognition phase. This can take the form of using domain-specific knowledge to make more accurate corrections and using an OCR engine optimized for handwriting recognition.

Finally, we think there is scope within this approach to add semantic information to the assignment of values to keys. Currently, the approach relies solely on positions to make assignments. However, we have seen from many multimodal approaches to document understanding that text is highly informative \cite{xu2020layoutlm, xu2021layoutlmv2,garncarek2002lambert,cooney2023unimodal, li2021structext, li2021structurallm}. LAMBERT uses word tokens alongside bounding boxes for KIE \cite{garncarek2002lambert}, and \citet{majumder2020representation} use the data-type of fields to enhance value retrieval. Further work is required to investigate ways in which text can be integrated into our KIE approach.
\section{Conclusions}

In this paper, we present an end-to-end document classification and KIE technique that uses a novel application of assignment optimization to extract key information from insurance claim forms. The system is end-to-end in that accurate document classification is required to select a specific template to be applied downstream to the KIE task. We experiment with several encoding methods for documents, and use cosine similarity to measure the distance between a form and a bank of templates. Key values are extracted from forms and assigned to a corresponding value position in a template using an optimization algorithm. The noisy scanned documents used to validate the approach require substantial preprocessing through realignment and scaling.

A mean f1 score of 0.94 indicates that this is a promising new approach to KIE from structured forms. An ablation study indicated preprocessing stages are essential to optimize performance of this approach. Our analysis of results obtained from different documents has suggested potential areas where this approach can be enhanced with further development. The approach is particularly suited to industrial applications in which large volumes of identical forms with different information require extraction of key information.

\bibliography{references}
\bibliographystyle{acl_natbib}
\label{sec:bibtex}

\end{document}